\documentclass[a4paper]{article}

\usepackage{ISCSLP2021}
\usepackage{mathtools}
\DeclareMathOperator*{\argmin}{argmin}
\title{The NPU System for the 2020 Personalized Voice Trigger Challenge}
\name{Jingyong Hou, Li Zhang, Yihui Fu, Qing Wang, Zhanheng Yang, Qijie Shao, Lei Xie}
\address{
  Audio, Speech and Language Processing Group (ASLP@NPU), \\
  School of Computer Science, Northwestern Polytechnical University, China}
\email{\{jyhou,lxie\}@nwpu-aslp.org}
\begin{document}
\maketitle
\begin{abstract}
This paper describes the system developed by the NPU team for the 2020 personalized voice trigger challenge. 
Our submitted system consists of two independently trained sub-systems: 
a small footprint keyword spotting (KWS) system and a speaker verification (SV) system. 
For the KWS system, a multi-scale dilated temporal convolutional (MDTC) network is proposed to detect wake-up word (WuW). 
For the SV system, we adopt ArcFace loss and supervised contrastive loss to optimize the speaker verification system. 
The KWS system predicts posterior probabilities of whether an audio utterance contains WuW and estimates the location of WuW at the same time. 
When the posterior probability of WuW reaches a predefined threshold, the identity information of triggered segment is determined by the subsequent SV system. 
On the evaluation dataset, our submitted system obtains a final score of 0.081 and 0.091 in the close talking and far-field tasks, respectively. 
\end{abstract}

\section{Introduction}
Keyword spotting (KWS) aims to detect pre-defined keyword(s) from audio and one important application of KWS is wake-up word (WuW) detection, which is usually used to trigger a speech interface in various devices such as smartphone, smart speaker and different kinds of IoT gadgets. 
For some personal devices, e.g. smartphone, smart watch, ear-buds, users usually do not want other people to wake up their devices. 
To build a personalized WuW system that can be only triggered by the device owner, a speaker verification (SV) module is usually used to perform authentication after the WuW detection. The SV system is to identify whether a test utterance match the speaker's enrollment utterance and accept or reject the identity claim of the speaker accordingly.


KWS and SV have been extensively studied (most independently) in the past. 
As a flagship event of ISCSLP, the 2020 personalized voice trigger challenge (PVTC2020) combines WuW and SV tasks together, providing a sizable dataset and a common testbed which is able to train and test a personalized voice trigger system. More details about the task setting, dataset and evaluation plan can be found in~\cite{jia20212020}. We have participated in the challenge and our submitted system ranks 2nd among all submitted systems in both close talking and far-field tasks, with a detection cost of 0.081 and 0.091 respectively\footnote{https://www.pvtc2020.org/leaderboard.html}.

\section{Methods}
\begin{figure}[tp]
  \centering
  \includegraphics[height=3.5cm]{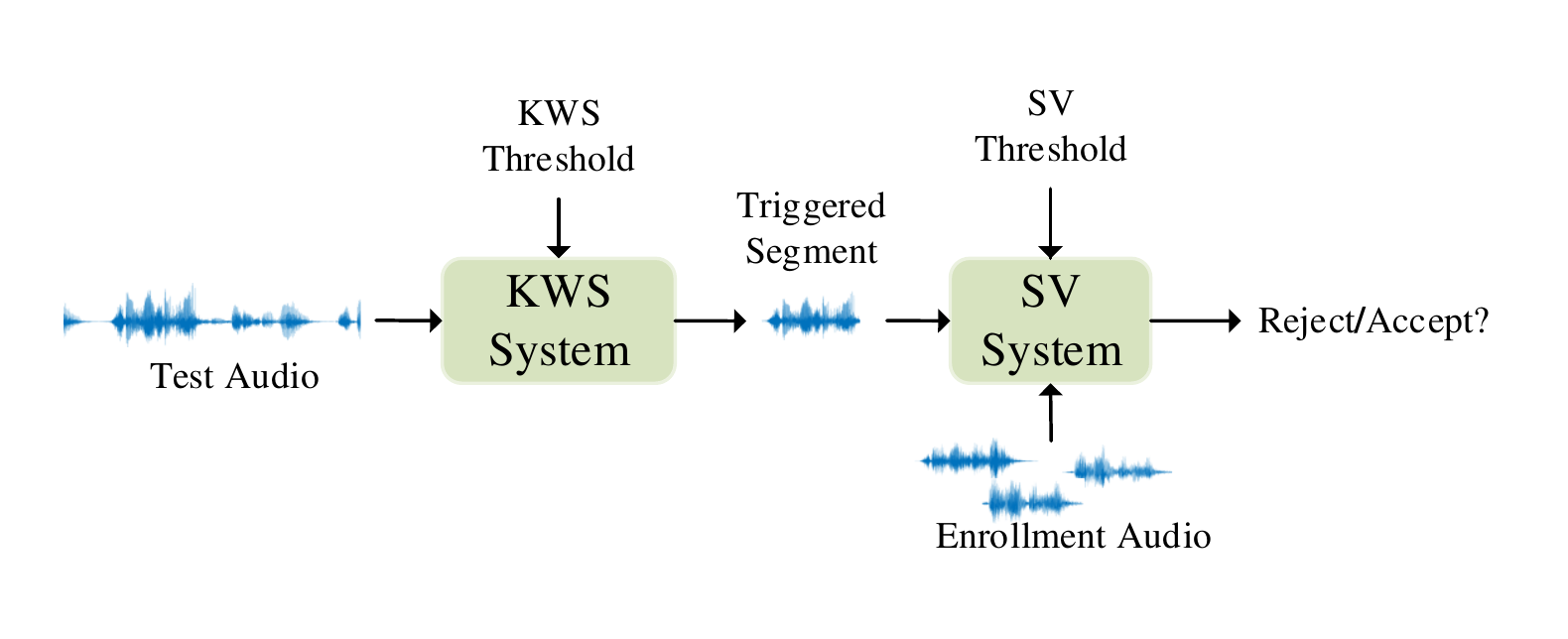}
  \caption{Framework of our proposed personalized voice trigger system.}
  \label{fig:framework}
\end{figure}

Our system consists of two independently trained sub-systems: 
a small footprint KWS system and an SV system, as shown in Figure~\ref{fig:framework}. 
The KWS system takes test audio as input and outputs triggered segment. 
After that, the triggered segment will be fed to the SV system. 
Then, the SV system decides whether the triggered segment is spoken by the enrolled target speaker. 

\subsection{KWS Sub-system}
\subsubsection{KWS Detector}
\begin{figure}[tp]
  \centering
  \includegraphics[height=6cm]{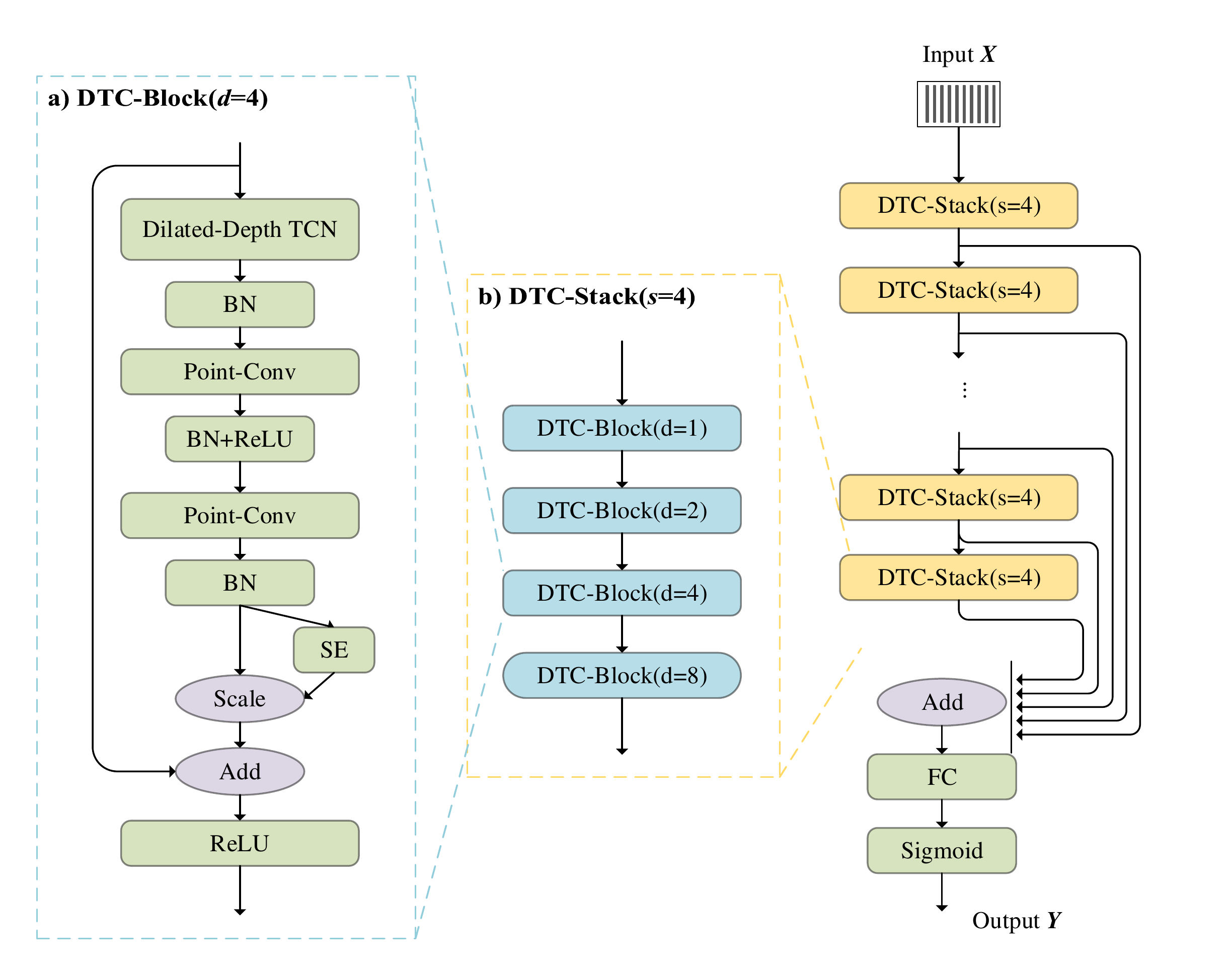}
  \caption{MDTC based KWS system.}
  \label{fig:PVTC_KWS}
\end{figure}
Our proposed end-to-end (E2E) KWS system is shown in Figure~\ref{fig:PVTC_KWS}. 
It takes $T$ frame mel-filter banks (80 banks per frame) $\mathbf{X}=(\mathbf{x_1},\mathbf{x_2}, ..., \mathbf{x_T})$ as input 
and outputs $T$ WuW posterior probabilities $\mathbf{Y}=(y_1,y_2, ..., y_T)$. 
For each time frame $t$, once its WuW posterior probability $y_t\ge\gamma$, wake-up word has judged as occurred. 
$\gamma \in (0, 1)$ is a threshold. 

To model the acoustic sequence in our keyword detector, a multi-scale dilated temporal convolutional (MDTC) network is proposed. 
A basic block, namely DTC block, is shown in Figure~\ref{fig:PVTC_KWS} (a). 
First, a dilated depthwise 1d-convolution network (Dilated-Depth TCN) is used to obtain temporal context with filter size is (5*1) and dilation rate can be set accordingly. 
Since simple depthwise 1d-convolution is used, the number of training parameters and computation cost can be reduced greatly. 
After the Dilated-Depth TCN, two layers of pointwise convolution (Point Conv) are used to integrate the features from different channels. 
We insert batch normalization (BN) and ReLU activation functions between different convolutional layers. 
In addition, we add a squeeze-and-exception (SE) module after the last Point Conv layer to learn the attention information between different channels. 
A residual connection between input and last ReLU activation function is also adopted to prevent gradient vanishing and gradient explosion. 
Four DTC blocks are stacked to form a DTC stack, as shown in Figure~\ref{fig:PVTC_KWS} (b). The dilation rates of four DTC blocks are set to 1, 2, 4 and 8 respectively. 
The receptive field of each DTC stack is 60 frames. 
In our submitted system, we use 4 DTC stacks as the feature extractor. 
Receptive field of the feature extractor is $4*60=240$ frames, which is big enough to model a WuW.
We extract feature maps from DTC stacks with different receptive fields and sum them up as the input of a keyword classifier. 
For the keyword classifier, a simple fully-connected layer followed by a sigmoid output layer is used to predict the posterior of WuW. 

\subsubsection{KWS Labeling and Loss}
For a positive training utterance, we select up-to 40 frames around middle frame of the WuW region as positive training samples and assign $1$ to them. 
Other frames in the positive training utterance are discarded as ambiguous and are not used in training. 
For negative training utterances, all frames are regarded as negative training samples and assigned to $0$. Our KWS system is thus modeled as a sequence binary classification problem. 
To train the model, binary cross entropy (BCE) loss is used:
\begin{equation}
  Loss(BCE) = -y_{i}^{*}\ln{y}_{i}-(1-y^{*})\ln(1-y_i),
\end{equation}
where $y_{i}^{*}\in\{0,1\}$ is the ground-truth class label for frame $i$, 
$y_i =M(\mathbf{x}_i;\theta) \in (0,1)$ is the WuW posterior predicted by the KWS model $M$ with parameter $\theta$.

\subsubsection{Data Augmentation}
We augment the original training set of the challenge, which is critical for our system to generalize to the development set as well as the evaluation set. 
For the original training data, the keyword always appears at the beginning of the utterance and rest of the utterance is the speech spoken by the same speaker. 
For the development set and the evaluation set, the keyword always appears at the end of the utterance, 
and there may exist speech from other speakers before the keyword. 
Only using the original training data to train our model, it is hard to generalize the model to the development and evaluation sets,  which results in a very low recall. 

 The positive keyword training set is composed as the following parts.
 \begin{itemize}
 \item[1)] The keyword segments in the training positive utterances;
 \item[2)] Randomly select non-keyword speech segments and pad them before the above keyword segments in 1); 
 \item[3)] Pad non-keyword speech segments both before and after the above keyword segments in 1);
 \end{itemize}

In addition, we also create more negative training utterances.
The specific strategy is to cut the positive utterance in 3) at the middle frame of the keyword into two segments which are subsequently used as negative training examples. This kind of negative examples can improve the generalization ability of the model too.

SpecAugment~\cite{Park2019} is also applied during training, which is first proposed for end-to-end (E2E) ASR to alleviate over-fitting and has recently proven to be effective in training E2E KWS system as well~\cite{hou2020small}. 
Specifically, we apply time as well as freqency masking during training. 
We randomly select $0-20$ consecutive frames and set all of their Mel-filter banks to zero, for time masking. 
For frequency masking, we randomly select $0-30$ consecutive dimensions of the 80 Mel-filter banks and set their values to zero for all frames of the utterance.

\subsubsection{Location Estimation}
As mentioned before, the keywords always appear at the end of positive utterances. Based on this, we do not explicitly predict the starting and ending frames of the keywords. Instead, in an utterance, we take the frame with the largest keyword posterior as the middle position of the keyword. We use the estimated middle position and the end frame of the keyword to estimate the starting frame of the keyword. Although this trick can not be applied to real applications, it is effective for the specific condition of this challenge. Note that there are several previous studies that explicitly model the location of keyword in a positive utterance~\cite{hou2019region,segal2019speechyolo,2019Simultaneous,jose2020accurate}.

\subsection{SV Sub-system}

\subsubsection{Speaker Verification Model}
\label{sec:sv_model}
We use the similar model structure as the baseline~\cite{jia20212020}. 
Our SV system consists of a front-end feature extractor, a statistic pooling layer and a back-end classifier. 
ResNet34~\cite{he2016deep} with SE-block~\cite{hu2018squeeze} is used as the feature extractor. 
For back-end classifier, the ArcFace loss~\cite{deng2019arcface} (the Baseline system uses AM-Softmax) is used. 
Supervised contrastive loss~\cite{khosla2020supervised} is also adopted to further decrease the intra-class distance and increase the inter-class distance. 
We have trained two models. The two models have the same model structure except for the statistic pooling layer. 
One model uses attentive statistic pooling (ASP)~\cite{okabe2018attentive}, 
and the other uses the self-attentive pooling (SAP)~\cite{rahman2018attention}. 

\subsubsection{Training Corpus}
We choose SLR38, SLR47, SLR62, SLR82 and SLR33 from OpenSLR\footnote{https://openslr.org. Note that speakers with less than 10 utterances are excluded in model training} to pretrain our SV model. Data augmentation strategies in Kaldi~\cite{povey2011kaldi} are adopted duration training. 
The MUSAN~\cite{snyder2015musan} noise and room impulse response (RIR) databases from~\cite{allen1979image} are used for noisy speech simulation. Eighty dimensional Mel-filter bank features with 25ms window size and 10ms window shift are extracted as model inputs.

\section{Experimental Setups}
\subsection{KWS}
We randomly separate $10\%$ speakers' data in the original training set as a validation set, and the rest participates in the back propagation. Adam optimization is used with a learning rate of $0.002$ and a mini-batch size of $M=150$. 
After each epoch, we evaluate the loss on the validation set.
If there is no reduction in loss, the learning rate begins to decay by a factor of $0.7$.
After at least $15$ epochs of training, we stop the training if there is no further decrease in the loss on the validation set.

\subsection{SV}
To train the SV system, we keep the same training-validation data configuration as the KWS system. We first pretrain our base model using the selected data from OpenSLR. After that, we fine-tune the model twice using the training data. 
 The original training dataset is used to adapt the base model at first. Then, to obtain a text-dependent speaker verification system, the keyword segments (cut from the whole utterance) is used to finetune the model again. 
Stochastic gradient descent (SGD) optimizer is used for model training. 
During pretraining, the initial learning rate is 0.1, decayed 10 times every 5 epochs. 
After 30 epochs, the loss will converge to around 0.2. 
In the finetune stage, we freeze parameters of the pretrained model in the beginning and only train the final layer. When the loss reaches around 0.2, all the parameters are tuned in the training, until the loss converge to a steady state. 

\subsection{Evaluation Metric and Determination of Thresholds}
In this challenge, a special detection cost function $C_d$ defined by the organizers is used to evaluate a personalized voice trigger system. 
False rejection (FR) and false alarm (FA) are two typical detection errors in the task.
Detection cost function is a weighted sum of FA rate (FAR) and FR rate (FRR):
\begin{equation}
 C_{d}=FRR+\alpha*FAR,
\end{equation}
where, $\alpha$ is a factor used to adjust the cost of FAR and FRR, and it is constantly set as $19$~\cite{jia20212020}.
Usually, by tuning the detection threshold of the SV system, FAR and FRR will change and there is a tradeoff between the two.
If we know the detection label of a dataset, we can get $minC_{d}$, a minimum $C_{d}$ on the dataset:
\begin{equation}
minC_{d} = \argmin_{\delta} (FRR(\delta)+\alpha*FAR (\delta)),
\end{equation}
where $\delta$ is the threshold of the SV system.

We have two thresholds need to be determined --  $\gamma$ for the KWS system and $\delta$ for the SV system. 
On the development set, we find that a lower KWS FRR could make the $minC_{d}$ smaller.
To this end, we choose a smaller $\gamma=0.01$ to ensure that the KWS FRR is relatively lower.
On the development set, we traverse all the SV thresholds to get $\delta_1$ and $\delta_2$ which minimize the $C_{d}$ for close talking and far-field tasks. 
On the evaluation set, the SV threshold for each task is directly borrowed from that tuned for the development set. 

\section{Experimental Results}
\subsection{KWS Results}
\begin{figure}[t]
  \centering
  \includegraphics[height=6cm]{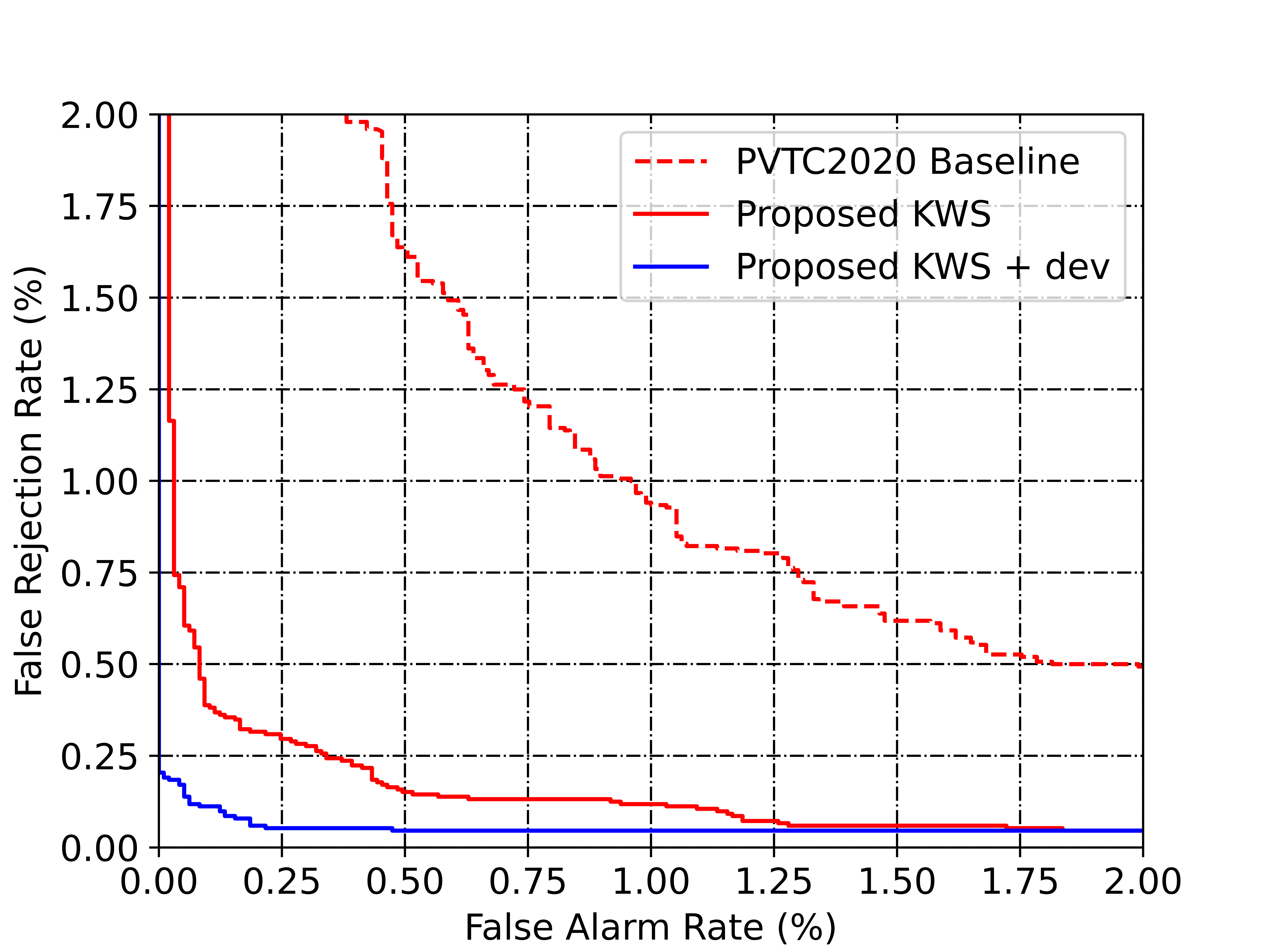}
  \caption{DET curves of different KWS systems on development set with close talking task.}
  \label{fig:KWS_results_task1}
\end{figure}

\begin{figure}[t]
  \centering
  \includegraphics[height=6cm]{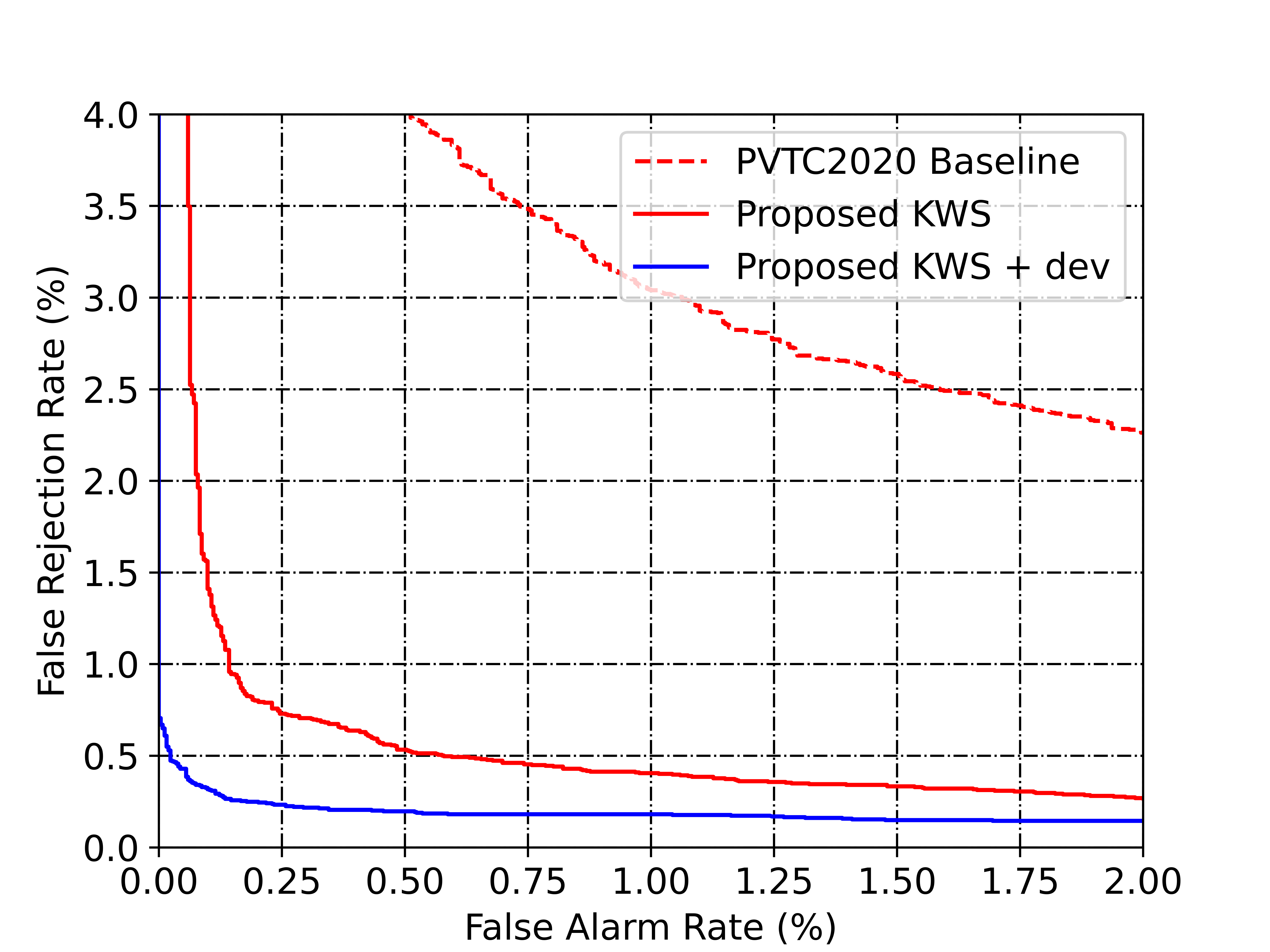}
  \caption{DET curves of different KWS systems on development set with far-field task.}
  \label{fig:KWS_results_task2}
\end{figure}
Figure~\ref{fig:KWS_results_task1} and~\ref{fig:KWS_results_task2} show DET curves for different KWS systems on the development set.
We reproduce the official KWS system (`PVTC2020 Baseline' in Figure~\ref{fig:KWS_results_task1} and~\ref{fig:KWS_results_task2}) according to the official source code. 
Compared to the PVTC2020 Baseline, our proposed KWS system achieves obviously much better DET curves in both close talking and far-field tasks.

In order to further improve the performance of our KWS system, the negative utterances of the development set are used to train our KWS model.
There are two reasons: 
1) we assume that the recording condition in development set is more matchable to that in the evaluation set, compared to the training set;
2) we want to use data from more speakers to improve the generalization ability of the model.

We do not have the label of evaluation set, so we still have to test the results on the development set even we have used it to train our KWS model.
As expected, using development negative utterances to train our KWS model could improve the performance by a big margin (`PVTC2020 Baseline + dev' in Figure~\ref{fig:KWS_results_task1} and~\ref{fig:KWS_results_task2}).

In the model training, $2/3$ of the training data is far-field data. However, comparing Figure 1 and Figure 2, we find that the performance on far-field task is much worse than that on close talking task. This indicates that the far-field task is more challenging than the close talking task.

\subsection{SV Results}
\begin{table}
  \caption{EERs of different SV systems on development dataset}
  \label{tab:tab_sv}
  \centering
  \begin{tabular}{c | c |c}
    \hline
      Model Name & Close talking data  &  Far-field data  \\
    \hline
    PVTC Baseline& 1.326   & 1.904    \\
    SV-ASP    & 0.961   & 1.564  \\
   SV-SAP    & 1.001   & 1.624  \\ 
    SV-Fusion & 0.821   & 1.423  \\
    \hline
    \end{tabular}
\end{table}
SV results on the development set are shown in Table~\ref{tab:tab_sv}. 
SV-ASP and SV-SAP are two models with different statistic pooling layers mentioned in Sec.~\ref{sec:sv_model}.
Here, equal error rate (EER) and is used as the SV evaluation metric.
Compared to the baseline SV system, our system (with ArcFace loss and supervised contrastive loss) obtains better ERRs on both tasks. We also notice that ASP and SAP have comparable EER performance.
Finally, simple score fusion on the two SV system brings further EER reduction.

\subsection{Results of the whole system}
\begin{table}
  \caption{The minimum detection costs $minC_d$ on development set and actual detection costs $C_d$ on evaluation set}
  \label{tab:tab1}
  \centering
  \begin{tabular}{c | c | c}
    \hline
    Dataset   &  Close talking data &	Far-field data\\
    \hline
    Development &  0.042 & 0.055 \\
    Evaluation  &  0.081 & 0.091 \\
    \hline
    \end{tabular}
\end{table}

Table~\ref{tab:tab1} shows personalized voice trigger performance of our system on different datasets and tasks. 
The values in the table are minimum detection costs $minC_d$ on the development set and actual detection costs $C_d$ on the evaluation set. 
On both development and evaluation sets, our system's performance in the far-field task is worse than that in in close talking task. This is consistent to the KWS results on the two tasks. 

Another observation is that on both tasks, performance on the development set is always better than that on the evaluation set. 
Our SV threshold determination method could be one reason to explain this.
Another possible reason is that, compared to the development set, there are more speakers in evaluation set and more speakers in the trials may make the SV task more challenging.

\subsection{Inference Efficiency}
An Intel (R) Xeon (R) E5-2620 v3 CPU with a main frequency of 2.4 GHz is used to evaluate the inference efficiency of our system. Our KWS model has around 180k parameters. 
On the evaluation set, the normalized real-time factor (RTF) of our KWS sub-system is 0.05. The SV sub-system does not need to process all the samples in the evaluation set. Instead, only the triggered segments will be sent to the SV system. To calculate the real-time factor of the SV part in the whole system, we compute the processing time of triggered segments on the evaluation set, and then divide it by the total duration of evaluation set. Our SV system's normalized RTF is 0.07.

\section{Conclusions}
Our personalized voice trigger system submitted to PVTC2020 is introduced in this paper. 
Our system consists of a KWS system and an SV system. 
The KWS system and the SV system are independently optimized but eventually cooperate well.
For the KWS system, a novel MDTC network is proposed and data augmentation strategies are used.
For the SV system, we revise the baseline system by using the ArcFace loss and the supervised contrastive loss, which is shown to be effective for performance gain. On the evaluation dataset, our submitted system obtains a final score of 0.081 and 0.091 in the close talking and far-field tasks, respectively. We will focus on joint optimization of KWS and SV systems in the future. We believe this will bring performance gain apparently. 

\bibliographystyle{IEEEtran}

\bibliography{mybib}

\end{document}